\documentclass[sigconf]{acmart}

\def\vec#1{\mathchoice{\mbox{$\displaystyle\bf #1$}}
{\mbox{$\textstyle\bf#1$}}
{\mbox{$\scriptstyle\bf#1$}}
{\mbox{$\scriptscriptstyle\bf#1$}}}
\def\v#1{\protect\vec #1}

\usepackage[linesnumbered, ruled, vlined]{algorithm2e}
\usepackage{graphicx}
\usepackage{url}
\usepackage{subcaption}
\usepackage{makecell}
\usepackage{etoolbox}
\AtBeginEnvironment{algorithm}

\AtBeginDocument{
\providecommand\BibTeX{{
 \normalfont B\kern-0.5em{\scshape i\kern-0.25em b}\kern-0.8em\TeX}}}

\copyrightyear{2024}
\acmYear{2024}
\setcopyright{licensedothergov}\acmConference[WDC '24]{The 3rd ACM Workshop on the Security Implications of Deepfakes and Cheapfakes}{July 1--5, 2024}{Singapore}
\acmBooktitle{The 3rd ACM Workshop on the Security Implications of Deepfakes and Cheapfakes (WDC '24), July 1--5, 2024, Singapore}
\acmDOI{10.1145/3660354.3660355}
\acmISBN{979-8-4007-0649-3/24/07}

\begin{document}

\title{Honeyfile Camouflage: Hiding Fake Files in Plain Sight}

\author{Roelien C. Timmer}
\affiliation{
\institution{UNSW Sydney\\
CSIRO Data61\\Cyber Security CRC}
\country{Australia}
}
\email{r.timmer@unsw.edu.au}
\author{David Liebowitz}
\affiliation{
\institution{UNSW Sydney\\
Penten}
\country{Australia}
}
\email{david.liebowitz@penten.com}
\author{Surya Nepal}
\affiliation{
\institution{CSIRO Data61\\
Cyber Security CRC}
\country{Australia}
}
\email{surya.nepal@data61.csiro.au}
\author{Salil S. Kanhere}
\affiliation{
\institution{UNSW Sydney}
\country{Australia}
}
\email{salil.kanhere@unsw.edu.au}

\renewcommand{\shortauthors}{Timmer et al.}

\begin{abstract}
Honeyfiles are a particularly useful type of honeypot—fake files deployed to detect and infer information from malicious behaviour. This paper considers the challenge of naming honeyfiles so they are camouflaged when placed amongst real files in a file system. Based on cosine distances in semantic vector spaces, we develop two metrics for filename camouflage: one based on simple averaging and one on clustering with mixture fitting. We evaluate and compare the metrics, showing that both perform well on a publicly available GitHub software repository dataset.
\end{abstract}


\noindent \textbf{Honeyfile Camouflage: Hiding Fake Files in Plain Sight}\\

\noindent   Roelien C. Timmer, David Liebowitz, Surya Nepal, Salil S. Kanhere\\

This is a \textbf{camera-ready version} of the article accepted for publication at the \emph{3rd Workshop on the security implications of Deepfakes and Cheapfakes (WDC)} co-located with the \emph{ACM ASIACCS 2024}.\\

Please cite this pre-print version as follows.\\

\noindent {\fontfamily{qcr}\selectfont @inproceedings\{timmer2024honeyfile,\\
\noindent author = \{Timmer, Roelien C. and Liebowitz, David and Nepal, Surya and Kanhere, Salil S.\},\\
title = \{Honeyfile Camouflage: Hiding Fake Files in Plain Sight\},\\
year = \{2024\},\\
maintitle =  \{ACM ASIACCS 2024\},\\
booktitle = \{3rd Workshop on the security implications of Deepfakes and Cheapfakes (WDC)\},\\
pages = \{(to appear)\},\\
location = \{Singapore\},\\       
\}

}
\newpage

\maketitle

\section{Introduction}
Honeyfiles are fake files that can be deployed to detect intruders or malicious insiders~\cite{yuill2004honeyfiles}. Their mechanism of action is that of any honeypot~\cite{spitzner2003-1-honeypots}: well-designed honeypots are not attractive to legitimate users, who are typically familiar with the resources they need to get on with their jobs and are unlikely to interact with a honeypot. Any interaction with a honeypot during reconnaissance by an intruder, therefore, signals a likely breach or data theft attempt~\cite{spitzner2003honeypots, spitzner2003-1-honeypots}. Honeyfiles can contain beacons and trackers~\cite{bowen2009baiting}, and when designed with content similar to that of the real document assets they are protecting, can also suggest the interests and goals of the adversary through the file selection choices they make~\cite{liebowitz2021deception}. 

Quantifiable properties of honeyfiles are important, both as a means of measuring how well-designed a honeyfile is and to provide control over automatic generation. The Decoy Document Distributor (D$^3$)~\cite{bowen2009baiting}, for example, is a system that creates files containing embedded beacons or decoy credentials for endpoints that can be observed, as well as a host-based file access monitor. D$^3$ is designed around a set of core properties, asserting that a decoy document should be believable, enticing, conspicuous, detectable and differentiable and must be variable to maintain believability and not interfere with normal user access to documents. Whitham~\cite{whitham2014design} lists a similar set of requirements for automated generation, arguing that honeyfiles should be enticing, realistic, adaptive, minimise disruption, be scalable to provide 
protective coverage, minimise the risk of exposure to sensitive artefacts and copyright infringement and contain no
distinguishable characteristics.

These requirements lists are valuable when thinking about honeyfiles, but they tend to overlap in the sense that a change to any one quality of a honeyfile will likely affect others. The elements that are most significant to any given honeyfile deployment and how they could be quantified also depend on the environment and threat model. Consider, for example, honeyfiles deployed into a document repository or knowledge base accessed through an indexed search engine. A human intruder engaged in a search targeting topics of interest is likely to see snippets of text around the terms that are found by the search engine. Both the topics appearing in the honeyfiles, which determines how likely they are to be found by the search terms, and the realism of the snippets are important. Understanding the deployment environment and the adversary's perspective allows us to reason about which honeyfile properties to prioritise. We can also attempt to quantify those properties to control or monitor automated generation. 

Realistic content generation was, until quite recently, the primary challenge facing honeyfile creators. The development of Large Language Models (LLMs), image generators and multi-modal models has drastically reduced the effort required to do this automatically. Indeed, the concern has shifted to the threats such models pose and their capacity to produce convincing misinformation and disinformation at scale~\cite{weidinger2022taxonomy}. 

Enticement remains important as the means by which an intruder interacts with a honeyfile and is intimately tied to the content created by LLMs for text search and image generators for visual search. A content-dependent metric for enticement based on the topics present in a corpus of real files appears in ~\cite{timmer2022tsm}. It extracts key topics from the real files and compares them to honeyfile content in a vector semantic space that allows robust comparison to paraphrasing.

Given that realistic content is accessible and that we can quantify enticement, we consider what we believe to be the next most important characteristic of a honeyfile: \emph{camouflage}. Camouflage measures how well a honeyfile is hidden and how likely it is to be detectable as a fake artefact from its properties other than content. This file property could be considered an element of realism, but we want to distinguish it as a property that depends on the environment in which a honeyfile is deployed. Under this model, realism is an intrinsic property of a honeyfile that depends on the general type of documents it is mimicking, and camouflage is a function of the deployed environment. Like enticement, it depends on the real files amongst which the honeyfile is hiding. 

We are primarily interested in the realistic scenario where honeyfiles are created and distributed on a personal, shared or cloud filesystem and monitored for access, download or exfiltration. The malicious actor could even be a software agent programmed to seek files that contain specific content for exfiltration and able to examine the statistical properties of the data it encounters. The camouflage of a honeyfile in this scenario is a function of its filename and metadata. The honeyfile should not be identified as fake through obviously incongruous filenames or metadata or through quantitative analysis of the file names in one or more directories. Our objective is to generate well-camouflaged filenames that blend in. 

We note that camouflage conflicts with the D$^3$ property of being conspicuous. This reflects a different view of the adversary. D$^3$ decoys are aimed at human intruders navigating file systems for targets of opportunity, as reflected in the choice of filenames that reference passwords or credit cards. The idea of a camouflaged trap assumes a more sophisticated adversarial intent, deployed to find specific information and unlikely to be fooled by obvious fakes.

As a result, we may also require a different approach to triggering an alert if camouflaged files are subject to frequent accidental interaction. This may be as simple as setting the threshold for an alert to allow for false positives. We could also, for example, monitor an email boundary for documents of a specific type, such as contracts in the case of a law firm, that the organisation routinely sends. The detection of a honeyfile crossing this boundary has no benign explanation, and indicates a high probability of exfiltration, or at the very least a gross error by a staff member or process. Honeyfile interaction detection methods and parameters will depend on the nature and circumstances of an organisation. 
 
This paper presents two variants of a metric for camouflage that focuses on filenames, evaluating the distance from the average or closest cluster of local filenames in a semantic vector space. The camouflage metric depends on the distribution of cosine distances amongst the filenames in a semantic vector space, and we demonstrate the camouflage metrics on a dataset of software repository filesystems. 

Section~\ref{sec:related} describes the many, mostly heuristic, approaches to filename generation in the literature. Section~\ref{sec:vecspaces} provides a brief sketch of background material, word embeddings and clustering of directional data. The camouflage metrics appear in Section~\ref{sec:camo}, with results following in Section~\ref{sec:results}.

\section{Related Work}
\label{sec:related}
The earliest (and by far the most well-known) use of honeyfiles was Clifford Stoll's fabrication of a set of documents on the Lawrence Berkeley Laboratory computing system in the 1980s. Stoll, having detected and monitored an attacker who had broken into the system, wrote the documents to occupy and delay him while his location was traced~\cite{stoll1988stalking}. In his book \emph{The Cuckoo's Egg}~\cite{stoll2005cuckoo}, Stoll describes in detail how the documents were crafted to appeal to the interests of the intruder but neglect to mention how he named the files. Indeed, this vital detail is omitted even from the made-for-TV movie that followed the book's success.

Subsequent authors have been more accommodating, describing a range of approaches to naming honeyfiles, many aimed at making them conspicuous and enticing. 

Yuill \emph{et al.}~\cite{yuill2004honeyfiles} developed the first honeyfile intrusion detection system. In this paper, users of the honeyfile system are advised to name the files so that they will be unlikely to open a honeyfile by accident by using, for example, a name that would appear unusual only to the owner. Four types of files are considered of interest to intruders, with examples such as \emph{passwords.txt} and \emph{customer-accounts-system.pdf} given for those that contain information about access to other systems.

The paper introducing D$^3$~\cite{bowen2009baiting} describes an evaluation in which files were given conspicuous names, which included \emph{stolen passwords}, \emph{credit card}, \emph{private data} and \emph{Gmail account info}. Subsequent papers that use D$^3$ evaluate strategies to name and place honeyfiles. 

Ben Salem \emph{et al.}~\cite{salem2011decoy} evaluates the effectiveness of honeyfiles against masquerade attackers. In one experiment, test subjects obtained honeyfiles from D$^3$ and were given instructions to name the files so that they were enticing to the adversary but recognisable as a decoy by the subject. Amongst the conclusions reached by the study was the observation that such differentiability to the user was not sufficient to avoid false positives since some files were scanned by automatic processes, and this triggered an alert. 

A later paper that uses D$^3$~\cite{voris2013bait} proposes a strategy in which honeyfiles are placed in directories that have high numbers of office document-related extensions such as \emph{.docx} and \emph{.xls} as well as recently accessed documents. Honeyfiles are named with one of three methods that "blend them with existing documents in a directory", indicating a shift from making them conspicuous to using camouflage. The first takes an existing filename and appends the string \emph{-final} or \emph{-updated}, the second appends a date to an existing filename, and the third modifies a filename to use the most common delimiter in the directory. 

A patent on honeyfile generation~\cite{whitham2018_patent} addresses honeyfile names as a special case of a general method. The approach treats natural language, document structures and filenames in a directory as token sequences. Tokens are grouped and classified by methods that depend on the sequence type. Using a real sequence as a template, fake sequences are generated by sampling from the token classes. In the case of filenames, tokens are extracted from filenames by classifying sub-strings into separators, numbers and letters. Sequences of letters can, in turn, be classified as lower, upper or camel case and split into words if lower case. The sequence tokens are then aggregated over one or more directories of file names to yield bags of tokens. A real filename can then be sampled from a directory, and its token sequence is treated as a template, into which samples from corresponding bags of tokens are drawn to generate a honeyfile name. 

The creation and detection of whole file systems as components of honeypots have also been investigated. Rowe~\cite{rowe2005automatic} presents a system that examines a file system with 28 metrics to determine if the file system may be a honeypot.
These metrics include the average number of files and programs per directory, the length of an average filename, the average file size, and average modification times. Surveying available studies on the characteristics of real filesystems, Whitham~\cite{whitham2014towards} develops metrics for creating whole filesystems as honeypots.

HoneyCode~\cite{nguyen2021honeycode} generates fake software repositories that include code, file and directory names. HoneyCode is trained on source code repositories mined from GitHub and uses RNNs to generate file and directory names and source code. A modified Graph Recurrent Attention Network is trained to generate a directory structure. 

\section{Semantic Vector Spaces}
\label{sec:vecspaces}

This section discusses different aspects of semantic vector spaces: word embeddings, cosine distance, von Mises-Fisher Mixtures and silhouette scores. These concepts form the basis of the camouflage metrics we propose.

\paragraph{Word embeddings}
Word embeddings are a technique in natural language processing (NLP) that maps words or phrases from the vocabulary to vectors of real numbers. This approach facilitates a representation of words in a continuous vector space, capturing their semantic and syntactic similarities~\cite{mikolov2013efficient}. However, a challenge with traditional word embedding models like word2vec~\cite{mikolov2013efficient} or GloVe~\cite{pennington2014glove} is their handling of rare or atypical words, which may not be represented if they do not occur in the training corpus. File and directory names often contain strings that are not natural language words, so we prefer to work with an embedding model that can handle these cases. FastText~\cite{bojanowski2017enriching}, an extension of the word2vec model, addresses this issue by representing words as bags of character n-grams, allowing it to generate embeddings for tokens not seen during training. This feature makes FastText particularly adept at handling words it was not trained on, as it can construct their embeddings by combining the embeddings of its subword units, thereby ensuring even strings with few or no occurrences in the training data can be represented in the vector space.

\paragraph{Cosine Distance}
Cosine similarity distance is a method to measure how different or similar word embeddings are. This is calculated using the formula: 
\begin{equation}
d(\v x ,\v y) =
 1-cos(\theta) = 1 - \frac{\v x\cdot \v y}{{\|\v x\| \|\v y\|}}
 \label{eq:cosine_distance}
\end{equation}
where $\v x$ and $\v y$ are the vectors for two texts, and $\theta$ is the angle between them. A smaller angle (or a cosine distance close to 0) means the texts are very similar, while a larger angle (cosine distance approaches 2) indicates they are completely different.

\paragraph{Von Mises-Fisher Mixtures}
We wish to identify clusters of similar filename representations in semantic vector space. Common clustering methods for text, such as K-means or hierarchical clustering, rely on Euclidean distance to measure the similarity between these vector representations. However, when dealing with text, directional clustering is often more appropriate because the direction of the vectors (rather than their magnitude) captures semantic similarity more effectively. The von Mises-Fisher (vMF) mixture model is a method for directional clustering that assumes data points are distributed on a unit hypersphere, making it well-suited for clustering high-dimensional text data where directionality is significant~\cite{banerjee2005misesfisher}. The probability density function of the vMF is:
\begin{equation}
p(\mathbf{x}; \mu, \kappa) = C_d(\kappa) \cdot \exp(\kappa \mu^\top \mathbf{x}), \quad C_d(\kappa) = \frac{\kappa^{d/2-1}}{(2\pi)^{d/2}I_{d/2-1}(\kappa)}
\end{equation}
where x is a unit vector on the d-dimensional sphere (the data point being considered), $\mu$ is the mean direction vector, $\kappa$ is the concentration parameter, and $C_d(\kappa)$ is the normalisation constant that ensures the total probability integrates to 1.

\paragraph{Silhouette score for clustering}

The silhouette score is a metric used to evaluate the quality of clusters, measuring how similar an object is to its own cluster compared to other clusters~\cite{rousseeuw1987silhouettes}. We define $a(i)$ as the mean distance between point $i$ and the other data points in the same cluster:
\begin{equation}
a(i) = \frac{1}{|C_i| - 1} \sum_{j \in C_i, j \neq i} d(i, j)
\end{equation}
where $C_i$ is the set of points in the cluster and $d(i, j)$ is the distance between $i$ and the other points in $C_i$.

Then $b(i)$ is the mean distance between datapoint $i$ and the datapoints in neighbouring cluster $C_k$:
\begin{equation}
b(i) = \min_{k \neq i} \frac{1}{|C_k|} \sum_{j \in C_k} d(i, j)
\end{equation}
and the silhouette score for point $i$ is :
\begin{equation}
s(i) = \frac{b(i) - a(i)}{\max\{a(i), b(i)\}}
\label{eq:silhouette}
\end{equation}
with $-1\leq s(i)\leq 1$. A high silhouette score indicates that the point is well-matched to its cluster and poorly matched to neighbouring clusters, meaning that clusters are distinct. The average silhouette score for a dataset is used to select the optimal number of clusters: 
\begin{equation}
ms = \sum_{i=1}^{N} \frac{s(i)}{N} 
\label{eq:mean_silhouette}
\end{equation}
where $N$ is the total number of data points.

For this work, we modify the silhouette score by replacing the Euclidean distance with the cosine distance.

\section{The Camouflage Metric}
\label{sec:camo}

In this section, we present two metrics to quantify camouflage for honeyfile names, a simplistic metric and a metric based on clustering (Section~\ref{sec:metrics}). We discuss the differences between these metrics by discussing an example scenario (Section~\ref{sec:example}). 

\subsection{Metrics}\label{sec:metrics}
 
We calculate a camouflage metric for a generated honeyfile name $g$. We have a directory $\mathcal{D}$ containing $D$ filenames: $\mathcal{D} = \{d1,.., d_D\}$. We compute embeddings for honeyfile name $g$ and the repository $\mathcal{D}$ with fastText~\cite{bojanowski2017enriching}, resulting in vectors $\vec g$ and $\{\v d_i\}$. As discussed in Section~\ref{sec:vecspaces}, we use fastText as an embedding model as directory and file names often consist of a combination of rare words. We can now define camouflage in two versions, one of which compares 
$\vec g$ to the mean of $\{\v d_i\}$, and one which considers 
clusters of similar names in $\mathcal{D}$.

\paragraph{Simple Camouflage}\label{sec:simple}
The simple camouflage score calculates the cosine distance from 
$\vec g$ to the mean of $\{\v d_i\}$.

\paragraph{Cluster Camouflage }\label{sec:cluster}
The cluster camouflage score first fits a mixture of vMF distributions to $\{\v d_i\}$. We calculate the order of the mixture by fitting $k=2..8$ using a stochastic gradient descent method~\cite{vmfclustering2014gopal} and selecting the best mean silhouette score from equation ~\ref{eq:mean_silhouette}. The cluster camouflage score for $\v g$ is the cosine distance to the mean of the closest mixture component.

We will illustrate the metrics with an example below, and in the next section test the metrics on a GitHub repository dataset. 
We introduce a new filename into a local directory and compute the camouflage scores. To keep this evaluation independent of a honeyfile generation mechanism, we sample files from other repositories in the dataset to compare to the local files.

The implementation of the simple camouflage and cluster camouflage score can be found on GitHub\footnote{Link to the GitHub implementation of simple camouflage and cluster camouflage: \url{https://github.com/RoelTim/honey-camouflage-nlp}}.

\subsection{Example}\label{sec:example}
\begin{figure}
 \centering
 \includegraphics[width=\linewidth]{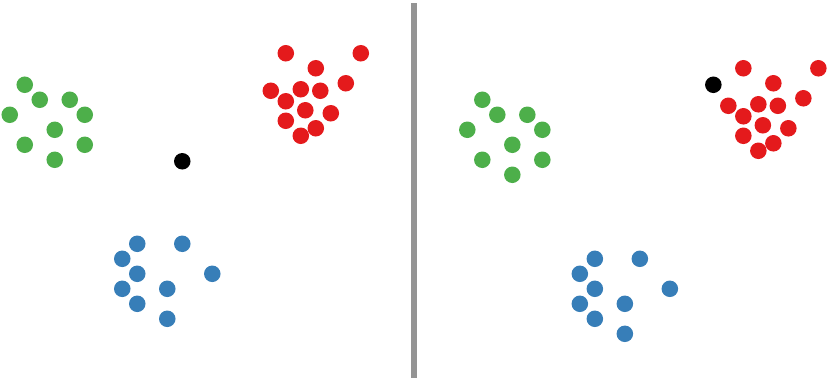}
 \caption{On the left side, a point (in black) is poorly camouflaged, whilst the point lies in the centre of all the points. On the right side, a point (in black) is well camouflaged as part of a cluster.}
 \label{fig:cluster_example}
\end{figure}

We visualise the camouflage measure concept in Figure~\ref{fig:cluster_example} with an analogous 2D representation (although note that semantic representations are directional). On the left, we see a novel point (shown in black) that is close to the centre of three clusters of points. It would be considered well camouflaged by the simple metric, even though it is not close to any of the clusters. Cluster camouflage, in contrast, would score the point on the left poorly camouflaged and the point on the right well camouflaged due to its proximity to one of the cluster centres. 

As a more realistic example, consider a directory containing the following filenames: \emph{`data1.xls', `data2.xls', `data3.xls', `data4.xls', `data5.xls',
 `regressions.r', `statistics.r', `evaluation.r', `testing.r',
 `report.pdf', `reportv1.pdf', `reportv2.pdf'}

We believe that \emph{data6.xls} would be well camouflaged in this directory. We calculate the cluster camouflage score for \emph{data6.xls}. Based on the silhouette score, the optimal number of clusters, $k*$, is 3. In Table~\ref{tab:example}, we rank all the files based on the shortest distance to the centroid for all three clusters.

The potential honeyfile name \emph{data6.xls} is the closest to the first centroid. Therefore, the cluster camouflage score is the distance to the first centroid: 0.060. For the simple camouflage score, we calculate the distance to the average vector, which is 0.120.

We can do the same for a filename that we think would not be a good honeyfile name for this type of directory \emph{wedding\_invites.xls}.
The cluster camouflage score is 0.173, and the simple camouflage score is 0.207.

\begin{table*}[ht]
\centering
\caption{Cosine distance of the local files to the centres of the three clusters. When underlined filename indicates the closest distance to that centre out of the three centres. }\label{tab:centroids}
\begin{tabular}{llllll}
\toprule
\textbf{Filename} & \textbf{\makecell{Distance to\\Centres \#1}} & \textbf{Filename} & \textbf{\makecell{Distance to\\Centres \#2}}& \textbf{Filename} & \textbf{\makecell{Distance to\\Centres \#3}} \\
\midrule
\underline{data4.xls}& 0.013& \underline{reportv2.pdf} & 0.153& \underline{statistics.r}& 0.157\\
\underline{data2.xls}& 0.023& \underline{report.pdf} & 0.220& \underline{regressions.r} & 0.161\\
\underline{data3.xls}& 0.026& \underline{reportv1.pdf} & 0.275& \underline{evaluation.r}& 0.161 \\
\underline{data5.xls}& 0.026& evaluation.r & 0.392& data4.xls& 0.262\\
\underline{data1.xls}& 0.033& data2.xls& 0.424& \underline{testing.r}& 0.271\\
statistics.r& 0.303& statistics.r & 0.429& data2.xls& 0.274\\
regressions.r & 0.376& data4.xls& 0.437& data5.xls& 0.302\\
evaluation.r& 0.438& regressions.r& 0.445& data3.xls& 0.311\\
reportv2.pdf& 0.471 & data3.xls& 0.447& data1.xls& 0.328\\
reportv1.pdf& 0.610 & data1.xls& 0.463& reportv2.pdf& 0.418\\
testing.r& 0.645& data5.xls& 0.467& reportv1.pdf& 0.595\\
report.pdf& 0.839 & testing.r& 0.468& report.pdf& 0.656\\
\bottomrule
\end{tabular}
\label{tab:example}
\end{table*}

\section{Results}
\label{sec:results}

This section presents the results of our experiments on GitHub repositories. We compare simple and cluster camouflage scores for local and sampled files, then analyse how these scores vary with directory size. 

\subsection{GitHub Repositories}
We conduct experiments on GitHub repositories to assess and compare the metrics. We use the Google BigQuery GitHub data set\footnote{Google BigQuery data set ID: \emph{bigquery-public-data.github\_repos}}. We select repositories consisting of at least 10 and a maximum of 500 items (directories and files). By only selecting a medium-sized repository, we avoid the likelihood of outliers or irrelevant data skewing the analysis results. Hereafter, we only select directories that consist of at least five items.

\subsection{Simple Camouflage}

\begin{figure}
 \centering
 \includegraphics[width=0.9\linewidth]{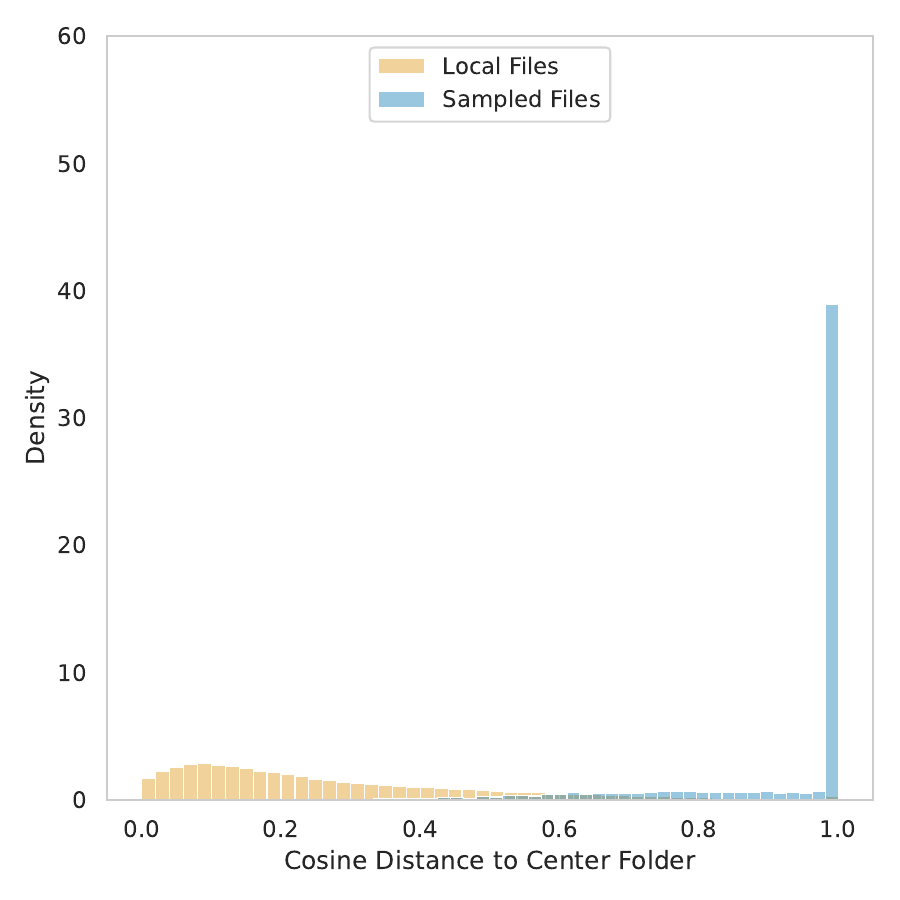}
 \caption{Distribution of the simple camouflage score for local and sampled files after normalisation for the scores per directory.}
 \label{fig:avg_folder_cosdist}
\end{figure}

We compare the simple camouflage score for local directory files and the sampled files from a different repository. We expect that the sampled files are unlikely to have names consistent with the local files since they are sampled from other repositories and thus have low camouflage scores. We experimented with 5,000 directories, and for each directory, we sampled a random file from a random repository, totalling 5,000 sampled files. Figure~\ref{fig:avg_folder_cosdist} shows the normalised simple camouflage score distribution for the local and sampled files.

The median simple camouflage score was 0.21 for the local files and 1.0 for the sampled files. The distribution of the scores between these two groups is significantly different, as shown by the Kolmogorov-Smirnov (KS) Test~\cite{smirnov1948table,Kolmogorov1933test}, which had a result of 0.85 and a p-value of 0.000.

\subsection{Cluster Camouflage}

\begin{figure}
 \centering
 \includegraphics[width=0.9\linewidth]{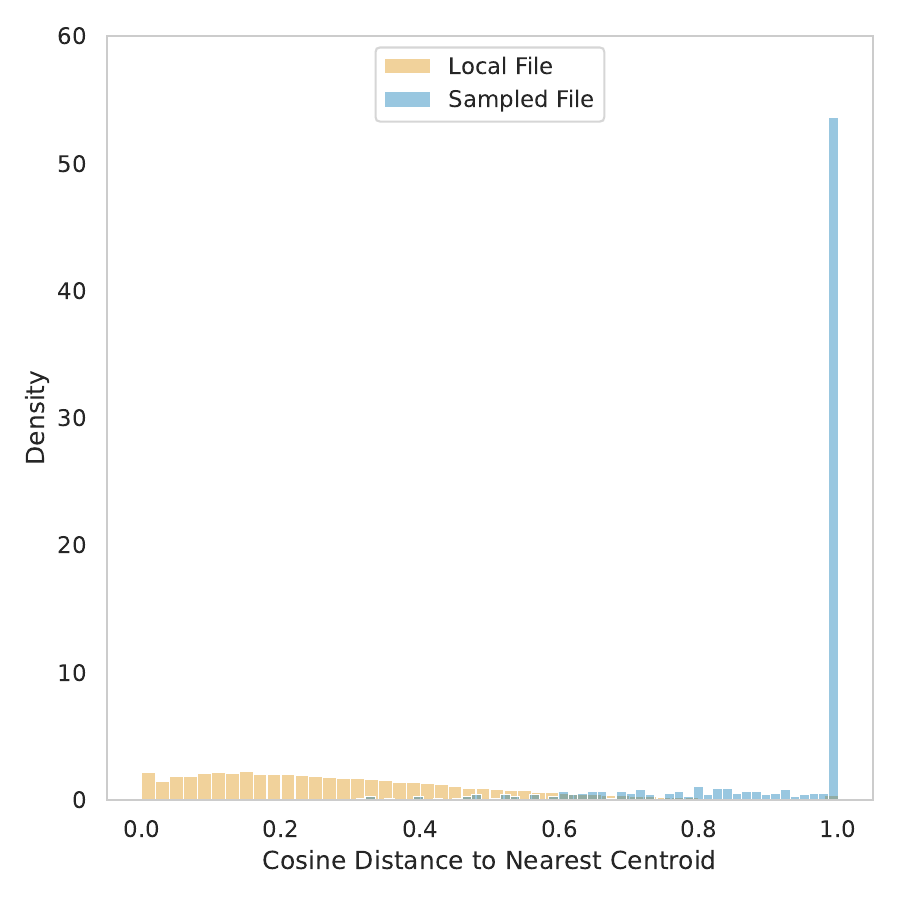}
 \caption{Distribution of the cluster camouflage score for local and sampled files after normalisation for the scores per directory.}
 \label{fig:vmf_folder_cosdist}
\end{figure}

The median cluster camouflage score is 0.23 for the local files and 1.0 for the sampled files. Figure~\ref{fig:vmf_folder_cosdist} shows the normalised cluster camouflage score distribution for the local and sampled files. The distribution of the scores between these two groups is significantly different, as shown by the KS Test, which had a test statistic of 0.88 and a p-value of 0.000.

We want to compare the results of the cluster camouflage score and simple camouflage score. The KS statistic is higher for the cluster camouflage score at 0.88, and 0.85 for the simple camouflage score. This difference suggests a greater disparity between the distributions of camouflage scores when using the cluster method compared to the simpler approach.

\subsection{Directory Sizes}
\begin{figure}
 \centering
 \includegraphics[width=.8\linewidth]{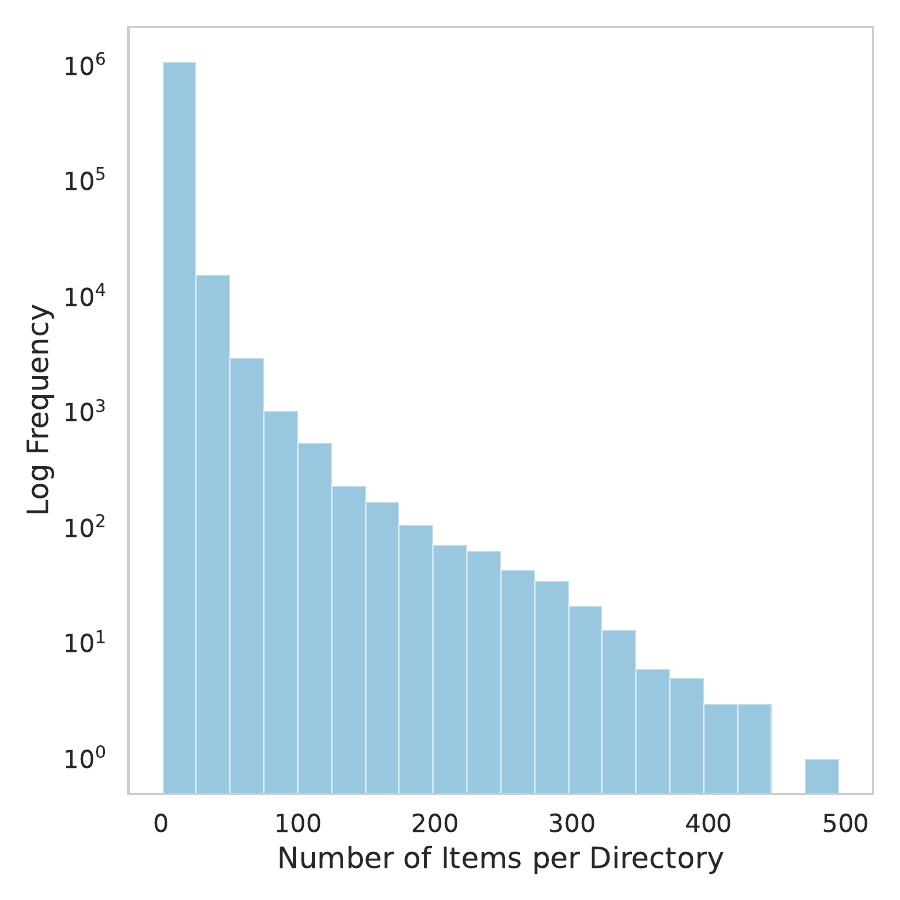}
 \caption{Log histogram of number of items per directory. }
 \label{fig:log_distribution_number_of_items_per_folder}
\end{figure}

We are also interested in whether the results for the simple camouflage and cluster camouflage scores depend on the directory size. We find that the number of items per directory follows a power law distribution (p-value=0.000) with $alpha=3.1$ and a cutoff at 30.00, meaning that a small number of directories contain a very large number of items, while a large number of directories contain a small number of items. Figure~\ref{fig:log_distribution_number_of_items_per_folder} shows the distribution of the number of items per directory where the y-axis is the log frequency. 

We compare small directories (5-10 items), medium directories (10-50 items), and large directories (50-500 items). Table~\ref{tab:small_medium_large_folders} shows the results of the KS Statistic of the distributions of the local and the sampled files. For the small directory, the KS statistics for the simple and cluster camouflage scores are very close, 0.88 and 0.87, respectively. For the medium directories, the difference is slightly bigger, 0.84 and 0.88. We find the most significant difference for the large directory, 0.90 and 0.84.

These results suggest that for large directories, the simple camouflage score would provide more insights as there is a greater distinction between the distributions of camouflage scores when using the simple camouflage method compared to the cluster approach. For the medium directories, we find similar results, but the difference is much smaller. For the medium directories, there is a slightly greater distinction between the distributions of camouflage scores when using the simple camouflage method compared to the cluster approach.

\begin{table}[htbp]
\centering
\caption{Camouflage score KS.}
\label{tab:camouflage_ks}
\begin{tabular}{cccc}
\toprule
\textbf{Directory Size} &\textbf{Score Type}& \textbf{KS Stat} & \textbf{p-value} \\
\midrule
Small &Simple Camouflage & 0.88&0.000 \\
&Cluster Camouflage&0.87 &0.000 \\
Medium &Simple Camouflage&0.84 &0.000\\
&Cluster Camouflage & 0.86& 0.000\\
Large &Simple Camouflage&0.90&0.000 \\
&Cluster Camouflage&0.84 &0.000\\
\bottomrule
\end{tabular}\label{tab:small_medium_large_folders}
\end{table}

\section{Conclusion}\label{sec:discussion}
The camouflage metrics performed well at identifying file names sampled from other repositories, suggesting that they can be used in practice to name deployed honeyfiles. 

We were surprised to observe that the two variants provided very similar scores in almost all cases. This is likely a result of the sampled filenames from other repositories being far enough from the local filenames that the distinction between clusters is negligible. This may or may not hold for other types of filenames. 

The simple camouflage score requires much less computation than the cluster score: it requires one averaging computation compared to K computations per iteration in mixture fitting.
While it is positive that the simpler computation yields an effective result, we are reluctant to discard the cluster-based approach without testing on other repository types. 
We limited ourselves to using GitHub repositories because we found no other publicly available large-scale filesystem datasets that include file names. We intend to test the camouflage metrics on other repository types and are particularly interested in corporate filesystems. 

In the future, we would like to analyse the camouflage scores for honeyfile names generated with LLMs. Generating filenames with LLMs could significantly enhance the efficacy of deception strategies in cybersecurity by making it more challenging for adversaries to distinguish between genuine and decoy files.

In closing, we note that creating a camouflage metric is challenging because of our uncertainty about the adversary. First, we do not know whether a software agent or a human is the infiltrator. Second, we do not know for sure what a human or a software agent infiltrator bases their decisions on. If a software agent is the infiltrator, the decision-making will likely be based on some pre-defined metrics. However, when a human is the infiltrator, the decision-making process might be more complex and harder to quantify. This asymmetric information situation is a challenge not only for designing a camouflage metric but also for every type of cyber deception metric.

\begin{acks}
The authors would like to thank UNSW, the Commonwealth of Australia, and the Cybersecurity Cooperative Research Centre Limited for their support of this work. The authors thank all the anonymous reviewers for their valuable feedback.
\end{acks}

\bibliographystyle{ACM-Reference-Format}
\bibliography{deception}

\end{document}